\documentclass[12pt,preprint]{aastex}

\newcommand{\Bband}{B}
\newcommand{\Vband}{V}
\newcommand{\BminusV}{({\Bband}{\rm -}{\Vband})}
\newcommand{\bminusv}{[{\Bband}{\rm -}{\Vband}]}

\newcommand{\EBV}{E\bminusv}
\newcommand{\Ebv}{E\BminusV}

\newcommand{\ebv}{$E\BminusV$}
\newcommand{\halpha}{H$\alpha$}

\newcommand{\hbeta}{H$\beta$}

\newcommand{\ssp}{\def\baselinestretch{1.0}\large\normalsize}
\newcommand{\gtrsi}{\mathrel{\hbox{\rlap{\hbox{\lower4pt\hbox{$\sim$}}}\hbox{$>$}}}}

\shorttitle{High Dust Polarization Efficiency} \shortauthors{Leonard et al.}
\begin{document}

\title{Evidence for Extremely High Dust Polarization Efficiency in NGC 3184}

\vspace{2cm}

\author{Douglas C. Leonard}
\affil{Five College Astronomy Department, University of Massachusetts, Amherst,
 MA 01003-9305; leonard@nova.astro.umass.edu}
\author{Alexei V. Filippenko, Ryan Chornock, and Weidong Li}
\affil{Department of Astronomy, University of California, Berkeley,
 California 94720-3411; (alex, chornock, weidong)@astro.berkeley.edu}

\vspace{1cm}

\begin{abstract}
Recent studies have found the Type II-plateau supernova (SN) 1999gi to be
highly polarized ($p_{\rm max} = 5.8 \%$, where $p_{\rm max}$ is the highest
degree of polarization measured in the optical bandpass; Leonard \& Filippenko
2001) and minimally reddened ($\EBV = 0.21 \pm 0.09$ mag; Leonard et al. 2002).
From multiple lines of evidence, including the convincing fit of a
``Serkowski'' interstellar polarization (ISP) curve to the continuum
polarization shape, we conclude that the bulk of the observed polarization is
likely due to dust along the line of sight (l-o-s), and is not intrinsic to
SN~1999gi.  We present new spectropolarimetric observations of four distant
Galactic stars close to the l-o-s to SN~1999gi (two are within $0.02^{\circ}$),
and find that all are null to within $0.2\%$, effectively eliminating Galactic
dust as the cause of the high polarization.  The high ISP coupled with the low
reddening implies an extraordinarily high polarization efficiency for the dust
along this l-o-s in NGC~3184: ${\rm ISP}/\EBV = 31^{+22}_{-9}\% {\rm\
mag}^{-1}$.  This is inconsistent with the empirical Galactic limit (${\rm
ISP}/\EBV < 9\% {\rm\ mag}^{-1}$), and represents the highest polarization
efficiency yet confirmed for a single sight line in either the Milky Way or an
external galaxy.

\end{abstract}

\medskip

\section{INTRODUCTION}
\label{sec:introduction}

It has long been understood that dust in the interstellar medium of the Milky
Way (MW) is responsible for the linear polarization of starlight from intrinsically
unpolarized Galactic stars (see, e.g., Greenberg 1968, and references therein).
The accepted model for the origin of the interstellar polarization (ISP) is
linear dichroism (directional extinction), which results from aspherical dust
grains aligned by some mechanism such that their optic axes have a preferred
direction.  By means of Mie theory computations (e.g., Greenberg 1968), the
extinction efficiency factors for a specific grain size, shape, and refractive
index may be calculated.  Typically, the dust grains are modeled as dielectric
cylinders with refractive index near $1.6$, the value for silicates. The basic
result is that an aspherical dust grain extinguishes slightly more light that
is propagating with electric vector parallel to its long axis than it does
light with electric vector perpendicular to its long axis. The resulting
polarization direction is therefore perpendicular to the direction of the
grain's alignment.  

The predicted degree of polarization produced by linear dichroism is strongly
wavelength dependent, with maximum polarization occurring when the
characteristic size of the grains (the cylinder's radius) is of the same
order as the wavelength of the light.  Unlike the behavior of dust extinction,
the expected spectral dependence of the ISP therefore rapidly decreases towards
either longer {\it or shorter} wavelengths away from the peak.  Such dust
polarization models have been quite successful at reproducing the
empirically derived ``Serkowski law'' ISP curve (Serkowski 1973; Wilking,
Lebofsky, \& Rieke 1982; Whittet et al. 1992) in the optical and near-infrared
bandpasses (see, e.g., Mathis 1986).

Naturally, the same dust that polarizes the light is expected to redden it as
well, a prediction observationally verified for MW stars.  The degree of
polarization produced for a given amount of extinction (or reddening) is
referred to as the ``polarization efficiency'' of the intervening dust grains.
The most efficient polarization medium conceivable is obtained by modeling the
dust grains as infinite cylinders (length $\gg$ radius) with diameters
comparable to the wavelength of the incident light, perfectly aligned with
their long axes parallel to one another and perpendicular to the line-of-sight
(l-o-s).  For such a model, Mie calculations place a theoretical upper limit on
the polarization efficiency of the grains due to directional extinction at
visual wavelengths of (Whittet 1992, and references therein)
\begin{equation}
\frac{\rm ISP}{\Ebv} <  R_V \times 13.82\%\ {\rm mag}^{-1} ,
\label{eq:3}
\end{equation}
\noindent where ${\rm ISP}$ is the maximum degree of interstellar polarization
measured in the optical band, \ebv\ is the reddening, and $R_V$ is the ratio of
total to selective extinction (e.g., Savage \& Mathis 1979).  This upper limit
on the polarization efficiency does not vary greatly with the refractive index
of the grains considered, and is therefore rather insensitive to uncertainty in
grain composition (Spitzer 1978, p. 175).  Adopting the canonical value of $R_V
= 3.1$ for MW dust leads to a maximum theoretical polarization efficiency along
typical lines of sight in the MW of ${\rm ISP}/\Ebv < 43\%\ {\rm
mag}^{-1}$.  This may be compared with the empirically derived upper bound
(Serkowski, Mathewson, \& Ford 1975) resulting from polarization studies of
reddened Galactic stars of
\begin{equation}
\frac{\rm ISP}{\Ebv} < 9.0\%\ {\rm mag}^{-1} .
\label{eq:4}
\end{equation}
\noindent The fact that the observed polarization efficiency in the MW is more
than a factor of 4 less than theory allows is generally taken as evidence that
the alignment of dust grains is not total (or has multiple preferred
orientations due to non-uniformity of the magnetic field along the l-o-s),
and/or that the grains are only moderately elongated particles (rather than
infinite cylinders) that may be irregularly shaped.

The empirical limit given by Equation~(\ref{eq:4}) is found to hold for
thousands of lines of sight to Galactic stars.  To be sure, there are several
stars in the comprehensive stellar polarization catalog by Heiles (2000) that
exceed this limit (indeed, several polarized stars have claimed values of $\Ebv
\leq 0.0$ mag, formally creating an infinite polarization efficiency), but the
colors and spectral types of the stars are not generally thought to be
determined with sufficient accuracy to claim a clear violation of the
inequality given by Equation~(\ref{eq:4}).

Somewhat surprisingly, there have been very few measures of the polarization
efficiency of dust in external galaxies.  By far the best way to determine the
polarization efficiency of dust in an external galaxy is to measure the ISP
along the l-o-s to a discrete source with a well-determined reddening.  Although
this has been a very rare exercise (see Hough 1996 for a review of the four
studies known at that time, and Rodrigues et al. 1997 for a subsequent study of
the polarization and reddening of stars in the Small Magellanic Cloud), the
results of the investigations that have been carried out are generally
consistent with the upper limit given by Equation~(\ref{eq:4}); while a few of
the Large Magellanic Cloud stars studied by Clayton, Martin, \& Thompson (1983)
do exceed the limit, the discrepancy has been attributed to incorrect spectral
classification or intrinsic colors (Hough 1996).

\begin{figure}
\ssp
\vskip -0.5in
\hskip -0.3in
\begin{center}
\rotatebox{0}{
 \scalebox{0.8}{
	\plotone{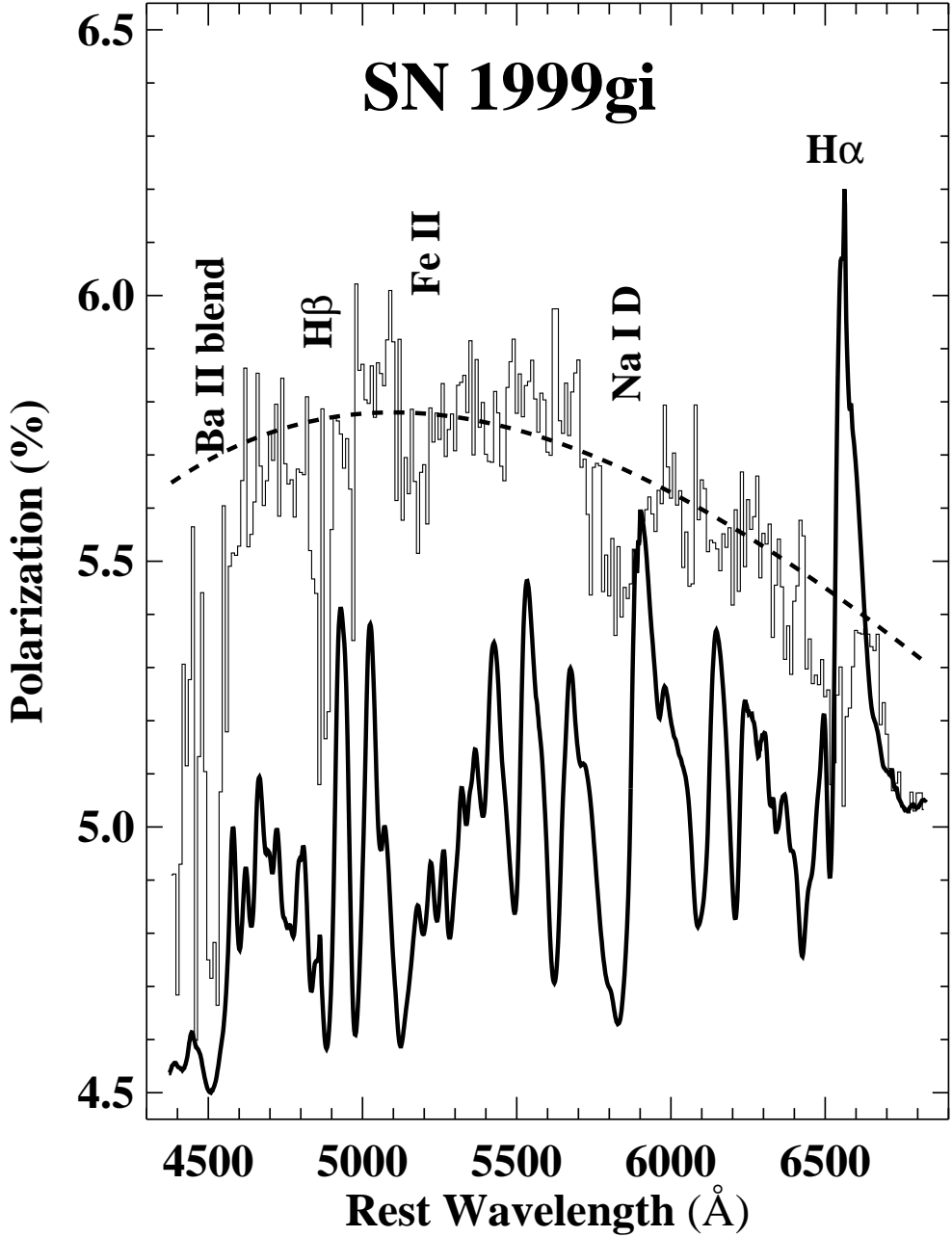}
		}
		}
\end{center}
\caption{Observed polarization (calculated as the rotated Stokes parameter; see
Leonard et al. 2001) of SN~1999gi ({\it thin line}) on 2000 Mar. 25 (107 days
after discovery; UT dates are used throughout this paper) compared with the
total flux spectrum from the same date ({\it thick line}), arbitrarily scaled
and offset for comparison of features.  A Serkowski ISP curve characterized by
$p_{\rm max} = 5.78\%, \theta = 154^\circ, {\rm\ and\ } \lambda_{\rm max} =
5100$ \AA\ is shown for comparison ({\it dashed line}; see
\S~\ref{sec:alternativestoisp}).  Prominent features in the spectropolarimetry
and total flux spectra are labeled.
\label{fig:1}}
\end{figure}

As part of a campaign to study the physical geometry of young supernovae (SNe;
Leonard, Filippenko, \& Matheson 2000b; Leonard et al. 2000a, 2001, 2002a),
Leonard \& Filippenko (2001; hereafter LF01) present a single epoch of optical
spectropolarimetry of the bright, Type II-plateau supernova (SN) 1999gi in
NGC~3184 during the late photospheric phase (107 days after discovery).  They
find an extraordinarily high degree of linear polarization, $p_{\rm max} =
5.8\%$, where $p_{\rm max}$ is the highest level of polarization observed in
the optical bandpass.  If intrinsic to SN~1999gi, such polarization is without
precedent for this type of event (see Wheeler 2000; Leonard et al. 2001; LF01).
It implies an enormous departure from spherical symmetry, since even the most
strongly aspherical theoretical models (axis ratio of 5-to-1, favorably viewed
edge-on) of H\"{o}flich (1991) fail to produce polarizations of more than $\sim
5\%$.  However, there is compelling evidence that much of the observed
polarization is, in fact, due to interstellar dust.  One clue pointing toward
an interstellar origin is that the polarization is not flat with wavelength,
but rather has a broad, asymmetric peak that gently decreases on either side
(Fig.~\ref{fig:1}).  This is unlike the wavelength-independent polarization
expected from an aspherical electron-scattering supernova atmosphere, but is
very similar to the polarization observed for reddened Galactic stars
(Serkowski 1973; Wilking et al. 1982; Whittet et al. 1992).  Since the same
dust that polarizes the SN light should also redden it, the natural expectation
is for significant reddening.  But, the Galactic foreground reddening being
only $\Ebv = 0.017$ mag (Schlegel, Finkbeiner, \& Davis 1998), LF01 contend
that nearly all of the dust along the l-o-s to SN~1999gi must reside in the
host galaxy, NGC~3184.

% **** Figure 1 ****

Leonard et al. (2002c; hereafter L02) present optical spectra and photometry of
SN~1999gi spanning the first six months after discovery as part of a study to
derive the distance to SN~1999gi through the expanding photosphere method.
Surprisingly, L02 conclude that SN~1999gi is only minimally reddened: Estimates
resulting from five independent techniques are all consistent with a hard upper
limit of $\Ebv < 0.45$ mag established by comparing the early-time color of
SN~1999gi with that of an infinitely hot blackbody, and yield a probable
reddening of $\Ebv = 0.21 \pm 0.09$ mag.  

In this paper, we further investigate the nature of the observed polarization
of SN~1999gi reported by LF01 in light of the recent evidence for low reddening
found by L02.  In \S~\ref{sec:observations} we present new spectropolarimetric
observations of four distant Galactic ``probe'' stars close to the l-o-s of
SN~1999gi in order to test the assertion by LF01 that the polarization is not
due to Galactic dust, and discuss the results in
\S~\ref{sec:galacticpolarization}.  In \S~\ref{sec:alternativestoisp} we
consider alternatives to ISP as the cause of the polarization, and
introduce a new technique to quantify the component of SN polarization that
lies in the direction perpendicular to the ISP direction in the Stokes
parameter $q$-$u$ plane.  In \S~\ref{sec:polarizationefficiencyofdustinngc3184}
we derive the polarization efficiency for dust in NGC~3184 along the l-o-s to
SN~1999gi.  We discuss the results in \S~\ref{sec:discussion} and summarize our
conclusions in \S~\ref{sec:conclusions}.

\section{Observations and Results}
\label{sec:observations}

\begin{figure}
\ssp
\begin{center}
\rotatebox{0}{
\scalebox{0.7}{
\plotone{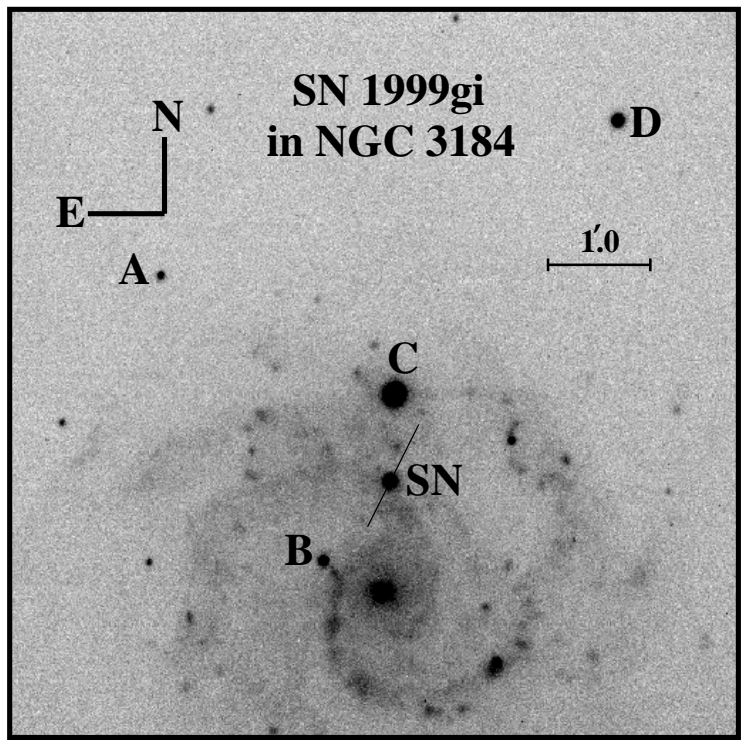}
}
}
\end{center}
\caption{$B$-band image of NGC 3184 taken on 1999 December 17 with the Katzman
Automatic Imaging Telescope (KAIT; Filippenko et al. 2001), with the four local
standards used by L02 marked.  Spectropolarimetry of Stars B and C (both within
$0.02^{\circ}$ of the l-o-s to SN~1999gi and distant enough to fully sample
Galactic dust) establishes that both are effectively null ($p_V < 0.2\%$),
essentially eliminating Galactic dust as the cause of the high polarization
found by LF01 for SN~1999gi.  The line segment through SN~1999gi ({\it SN}) at
P.A. = $154^\circ$ indicates the polarization angle of the SN measured by LF01.
\label{fig:2}}
\end{figure}

In order to investigate the contribution of Galactic ISP to the high
polarization reported by LF01 for SN~1999gi, we obtained spectropolarimetry of
four distant Galactic ``probe'' stars close to the l-o-s of SN~1999gi: BD
$+42^\circ2113$, HD 89572, and two ``local standard stars'' used by L02 to
calibrate the photometry (labeled {\it B} and {\it C} in Fig.~\ref{fig:2}).
The data were obtained on 2001 October 25 (BD $+42^\circ2113$ and HD 89572) and
2002 January 8 (Stars B and C), using the Kast double spectrograph (Miller \&
Stone 1993) with polarimeter at the Cassegrain focus of the Shane 3-m telescope
at Lick Observatory.  The data were reduced according to the methods detailed
by Leonard et al. (2001), and the results are given in Table~1.

\begin{deluxetable}{lccccccc}
\renewcommand{\arraystretch}{.6}
\tabletypesize{\scriptsize}
\tablenum{1}
\tablewidth{450pt}
\tablecaption{Probes of Galactic Dust Polarization Near the Line of Sight to 
SN 1999gi}
\tablehead{\colhead{Star} &
\colhead{Spectral Type\tablenotemark{a}} &
\colhead{$m_V - A_V$\tablenotemark{b}} &
\colhead{$M_V$\tablenotemark{c}}  &
\colhead{Distance\tablenotemark{d}} &
\colhead{Angular Separation\tablenotemark{e}} &
\colhead{$p_V$} &
\colhead{$\theta_V$} \\
\colhead{} &
\colhead{} &
\colhead{(mag)} &
\colhead{(mag)} &
\colhead{(pc)} &
\colhead{(deg)} &
\colhead{(percent)} &
\colhead{(deg)} }

\startdata

BD +$42^\circ2113$  & Am        & 8.83  & 0.6 & $>440 \pm 100$ & 0.52 & $0.07 \pm 0.01$ &
  $125 \pm 3$ \\
HD 89572            & A0        & 6.72  & 0.6 & $>170 \pm 40\ $ & 0.55 & $0.03 \pm 0.01$ &
  $98 \pm 8$  \\
Star B              & $\sim$ G6 & 15.41 & 5.5 & $>960 \pm 220$ & 0.02 & $0.19 \pm 0.07$ &
  $73 \pm 9$  \\
Star C              & $\sim$ G0 & 11.92 & 4.5 & $>300 \pm 70\ $ & 0.02 & $0.10 \pm 0.02$ &
  $111 \pm 5$ \\

\enddata 

\tablecomments{Stars B and C are the local standards labeled in Figure~2.  The
uncertainty in the polarization values is statistical only, and does not
incorporate the $0.1 - 0.2\%$ systematic uncertainty that may exist in
polarimetry measurements made with the Lick 3 m polarimeter (see \S~2).  For
low polarization values, the relatively large systematic uncertainty makes
$\theta_V$ much more uncertain than the statistical error alone implies.  Note
that $p_V$ and $\theta_V$ represent the debiased, flux-weighted averages of the
polarization and polarization P.A., respectively, over the wavelength range
$5050 - 5950$ \AA\ (see Leonard et al. 2001).}

\tablenotetext{a}{Taken from Simbad for BD +$42^\circ2113$ and HD 89572,
and estimated from colors and spectral features for Star B and Star C;
see text for details.}

\tablenotetext{b}{$A_V = 0.05 \pm 0.01$ mag (Schlegel et al. 1998) adopted for
all stars; $m_V$ taken from Simbad for BD +$42^\circ2113$ and HD 89572, and
from L02 (Table~1) for Stars B and C.  }

\tablenotetext{c}{For luminosity class V dwarfs, from Binney \& Merrifield
(1998, p. 107).  For BD +$42^\circ2113$, the $M_V$ for an A0 star is quoted,
since this classification is most consistent with its observed $B-V$ color. }

\tablenotetext{d}{Estimated from spectroscopic parallax with the assumption
that all stars are luminosity class V dwarfs; this assumption results in lower
bounds on the inferred distances.  The quoted uncertainty is derived by
assuming a $0.04$ mag uncertainty for ($m_V - A_V$) and a $0.5$ mag uncertainty
for $M_V$ (a rather liberal estimate, arising from the difficulty inherent in
exact spectral classification).}

\tablenotetext{e}{Between the probe star and SN~1999gi.  }

\end{deluxetable}

\clearpage

% **** Figure 2 ****

To determine the polarization angle (P.A.) offset between the half-wave plate
and the sky coordinate system on 2001 October 25, we observed the polarized
standard star BD $+59^\circ389$ and set its $V$-band polarization position
angle (i.e., $\theta_V$, where the $V$ subscript denotes the debiased,
flux-weighted average over the wavelength range $5050-5950$ \AA; see Leonard et
al. 2001) equal to $98.09^{\circ}$, the value cataloged by Schmidt, Elston, \&
Lupie (1992).  To check the precision of the P.A. offset we also observed the
polarized standard star HD 19820 on this night, and found its polarization
angle to agree to within $0.2^{\circ}$ of the value measured by Schmidt et
al. (1992).  The observed null standard, BD~$+32^\circ3739$ (Schmidt et
al. 1992), was found to be null to within $0.1\%$.

To determine the P.A. offset on 2002 January 8, we observed HD~58624 and set
$\theta_V$ equal to $26.9^{\circ}$ (Mathewson \& Ford 1970).  To check the
precision of the P.A. offset on this night we also observed the polarized
standard stars HD~43384 and HD~245310.  For HD~43384, $\theta_V$ was measured
to be $170.8^{\circ}$, which agrees to within $\pm 0.3^{\circ}$ of the value
cataloged by Mathewson \& Ford (1970).  However, the $V$-band polarization
angle of HD~245310 was found to be $148.0^{\circ}$, which disagrees by
$2^{\circ}$ with the value ($\theta_V = 146^{\circ}$) reported by Hiltner
(1956).  While a $2^{\circ}$ disagreement is not very alarming (and, in fact,
is not that uncommon when comparing with older catalogs), it is also possible
that a small amount of instrumental polarization existed on this night: the
observed null standard, HD~57702, was measured to have $p_V = 0.17\%$, which is
$0.13\%$ higher than the value reported by Mathewson \& Ford (1970).

The slight polarization for the null standard on 2002 January 8 could also be
explained by systematic uncertainty.  From repeated observations of a single
standard star, Leonard et al. (2001) estimate the systematic uncertainty of the
Lick polarimeter to be $\gtrsim 0.05\%$ in a Stokes parameter, or roughly $0.1
\%$ in the determination of $p_V$.  In addition, Leonard et al. (2001) show
that the uncertainty rises for objects with complex backgrounds, such as SNe in
dusty galaxies or Galactic stars superposed on such backgrounds.  Repeated
observations of SN~1999em, in fact, show disagreements in a single Stokes
parameter of up to 0.2\%.  Given these uncertainties, we shall consider any
measured polarization that is $< 0.2\%$ to be effectively null.  Fortunately,
although uncertainty at this level certainly places a cautionary note on the
interpretation of very low polarization measurements, it does not significantly
hamper the interpretation of large polarization values, such as that measured
for SN~1999gi.

\begin{figure}
\ssp
\begin{center}
\rotatebox{0}{
\scalebox{0.7}{
\plotone{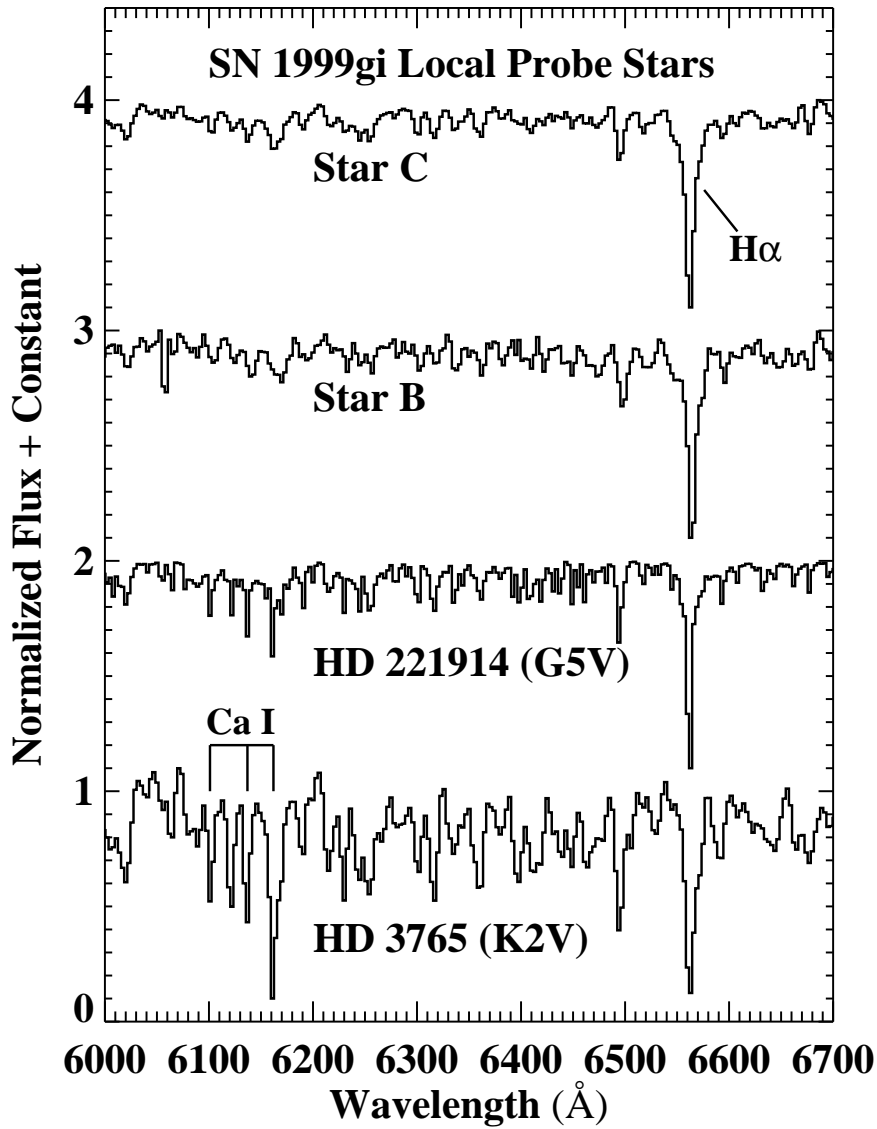}
}
}
\end{center}
\caption{Optical flux spectra of the two local dust probe stars discussed in
the text and labeled in Figure~\ref{fig:2}, with two additional spectra shown for
comparison of features.  All spectra have been normalized so that the \halpha\
line has the same depth in each spectrum.  The strength of the neutral metal
lines (e.g., the \ion{Ca}{1} lines indicated) in stars of type K2V (HD 3765)
and later effectively restricts the classification of the two probe stars to
earlier types.  This is consistent with the color-based classification of the
two stars of G0 (Star C) and G6 (Star B).  Note that the resolution of the two
comparison spectra ($\sim 2.5$ \AA) is somewhat higher than that of the local 
probe stars ($\sim 6$ \AA). The comparison spectra were taken using the 
Keck II 10-m telescope with the Low Resolution Imaging Spectrometer (Oke et 
al. 1995) as part of another program; see, e.g., Filippenko et al. (1999).
\label{fig:3}}
\end{figure}

In order to serve as good probes of interstellar dust polarization, it is
clearly necessary to observe stars that lie significantly above (or below) the
Galactic plane and close to the l-o-s of the SN.  Applying the criterion of
Tran (1995) that a good MW ``probe'' star should be more than 150 pc from the
Galactic plane, the Galactic latitude of SN~1999gi ($b \approx 55^{\circ}$)
necessitates using stars that are at least 183 pc away, a criterion that all
but one of our probe stars (HD 89572, $d = 170 \pm 40$ pc) definitely meets
(see Table~1).  We estimated the probe star distances through spectroscopic
parallax, with the spectral types and apparent magnitudes for BD
$+42^\circ2113$ and HD 89572 taken from the Simbad Astronomical
Database.\footnote{\url{http:simbad.u-strasbg.fr/Simbad}} To determine the
spectral types of the two ``anonymous'' local standard stars we compared their
observed colors, corrected for a Galactic reddening of $\Ebv = 0.017$ mag
(Schlegel et al. 1998; $\bminusv = 0.69 {\rm\ and\ } 0.57$ mag for Stars B and
C, respectively), with the values tabulated by Binney \& Merrifield (1998,
p. 107) for luminosity class V dwarfs.  (Note that if either of the probe stars
is in fact more intrinsically luminous than a class V dwarf, it would serve to
increase the inferred distance, which only enhances their efficacy as probe
stars).  We also verified this approximate classification through examination
of their spectra (Fig.~\ref{fig:3}).  With the possible exception of HD 89572,
then, all of the probe stars lie considerably farther away than the Tran (1995)
criterion requires, and are sufficiently close to the l-o-s to SN~1999gi to
serve as good probes of the Galactic ISP.

% **** Figure 3 ****

\section{Analysis}
\label{sec:analysis}
\subsection{Galactic Polarization}
\label{sec:galacticpolarization}

From Table~1 we see that all of the probe stars are effectively null,
confirming the suspicion of LF01 that the cause of the high polarization is not
local to the MW, but rather resides in the host galaxy, NGC 3184.  This is also
consistent with the low Galactic reddening ($\EBV\ = 0.017$ mag) reported by
the dust maps of Schlegel et al. (1998).  Since Stars B and C are located on
either side of SN~1999gi (Fig.~\ref{fig:2}) and are only $\sim 0.04^{\circ}$
away from each other, this constrains the possible radius of an undetected
polarizing spherical Galactic dust blob (or one dimension of a filamentary
structure) that covers only SN~1999gi and neither probe star to values less than
$\sim 0.13$ pc, assuming that the undetected blob resides within 150 pc of the
Galactic plane.  While such tiny concentrations of dust may be possible (e.g.,
the outer edge of a Bok globule, or a thin filamentary dust structure), the
chance for such an alignment is certainly quite remote.  We therefore feel
confident in asserting that Galactic dust is not responsible for the
polarization of SN~1999gi.  
\subsection{Alternatives to ISP}
\label{sec:alternativestoisp}

Before concluding that the high polarization of SN~1999gi is due to ISP from
dust in NGC~3184, however, we consider three alternatives. \\ 

\indent 1. {\it The polarization is due to newly formed dust in the expanding
SN ejecta.}  It is now generally accepted that dust formation is a natural
process in evolved SN atmospheres (e.g., Gerardy et al. 2000).  However, there
are several objections to this proposal for SN~1999gi at the epoch studied
here, about 111 days after the explosion (L02).  First, the temperature of the
ejecta at the observed phase, including the inner regions in which the dust is
thought to ultimately form (e.g., Wooden 1997; Clayton, Deneault, \& Meyer
2001), is greatly in excess of the $\sim 2000$ K evaporation temperature of
iron-rich grains (Wooden 1997), making it a very hostile environment for dust
survival (Rank et al. 1988; Dwek 1991; Wooden et al. 1993).  Indeed, the
detailed study of SN~1987A by Wooden (1997) shows that the earliest possible
epoch for dust to have formed in that event is about 350 days after explosion.
In addition, the spectral and photometric evolution of SN~1999gi do not exhibit
any of the telltale optical signatures of dust formation in the time leading up
to the spectropolarimetric observation, such as an accompanying drop in optical
luminosity or the onset of asymmetric line profiles (e.g., SN~1998S; Leonard et
al. 2000).  Dust formation in the ejecta of SN~1999gi can therefore be
confidently eliminated as a possibility at this comparatively early
evolutionary phase.

\indent 2. {\it The polarization is produced by dust {\it reflection} by one or
more off-center dust blobs in NGC~3184 near to the l-o-s of SN~1999gi.}
Reflection by dust is highly polarizing, and certainly could produce the
average level of polarization seen in SN~1999gi.  However, reproducing the
observed Serkowski-law dependence, while surely possible given all conceivable
dust-blob orientations and distances, would certainly require some fine-tuning
of the model since simple reflection (i.e., single scattering) off grains of
typical size produces polarization that increases at blue wavelengths due to
the greater scattering efficiency there (e.g., Miller \& Goodrich 1990; Miller,
Goodrich, \& Mathews 1991; see Kartje 1995 for a discussion of how multiple
scatters could alter this).  While computing a detailed light-echo model
(e.g., Wang \& Wheeler 1996) is beyond the scope of this study, we consider the
possibility of dust reflection mimicking a Serkowski-law ISP curve to be quite
remote.

\indent 3. {\it The observed polarization is intrinsic to SN~1999gi itself.}
It is clear that SN~1999gi has some intrinsic polarization.  The distinct
spectropolarimetric features seen in Figure~\ref{fig:1} correspond directly to
strong features in the total flux spectrum; such sharp polarization modulations
are not characteristic of ISP and must belong to the SN.  There are, however,
several reasons to believe that the intrinsic SN polarization component is
relatively minor compared with the ISP.

To start with, the strength of the line features relative to the continuum
polarization is quite small.  The basic paradigm used to interpret polarization
changes across P-Cygni line features is that polarization peaks are naturally
associated with absorption minima due to selective blocking of
forward-scattered (and hence less polarized) light in P-Cygni absorption
troughs and polarization minima are associated with emission peaks due to the
dilution of polarized continuum light with unpolarized line emission (e.g.,
McCall 1984; Jeffery 1991a; LF01; Leonard et al. 2002a).  In fact, the
polarization level in the emission components of strong lines (e.g., \halpha,
\ion{Ca}{2} infrared triplet) has been used in several previous studies to
define the ISP level (e.g., Trammell, Hines, \& Wheeler 1993; Leonard et
al. 2001, 2002a; Kawabata et al. 2002).  For SN~1999gi, the lack of any
strong polarization change across the \halpha\ emission component suggests that
the ISP level is not significantly different from the observed polarization.
In the absorption troughs, we see modest modulations at the locations of
\ion{Ba}{2} $\lambda 4554$, \hbeta, \ion{Na}{1} D, and possibly \ion{Fe}{2}
$\lambda 5169$ as well (Fig.~\ref{fig:1}).  Curiously, the polarization changes
in the troughs are observed to be {\it decreases} rather than the expected
increases; we shall discuss the implications of this unusual behavior later in
this section.  For now, we note that LF01 find the strength of the modulations
to be consistent with other SNe~II-P having low intrinsic continuum
polarization, and conclude that they are consistent with an intrinsic continuum
polarization as low as $p_{\rm SN} \approx 0.3\%$ for SN~1999gi.

Another argument against high intrinsic SN polarization is that the continuum
polarization shape is not flat with wavelength, as would be expected for an
aspherical electron-scattering atmosphere, but rather approximates a Serkowski
law ISP curve.  In Figure~\ref{fig:1} we show one possible Serkowski law fit to
the data, determined by finding the best agreement (lowest $\chi^2$) between a
Serkowski law ISP curve (Whittet et al. 1992) and the observed
spectropolarimetry in wavelength regions not obviously affected by line-trough
features (we chose to fit the regions 4600--4810~\AA, 5000--5100~\AA,
5230--5670~\AA, and 6070--6460~\AA).  The derived Serkowski law ISP curve shown
in Figure~\ref{fig:1} is certainly not a unique solution: the spectral regions
that were fit were chosen somewhat arbitrarily (indeed, the ``best'' fit
presented in Fig.~\ref{fig:1} is slightly different from that derived by LF01,
due largely to the different spectral regions that were fit) and {\it some}
intrinsic SN polarization definitely exists, which makes the true ISP level and
spectral shape slightly different from that to which we fit.  Nonetheless, the
good qualitative agreement between the overall shape of the observed
polarization and a Serkowski law ISP curve, and its incompatibility with
wavelength-independent polarization, is quite suggestive that ISP, and not
intrinsic SN polarization, dominates the observed polarization.

% **** Figure 4 ****

\begin{figure}
\ssp
\begin{center}
\rotatebox{0}{
\scalebox{0.7}{
\plotone{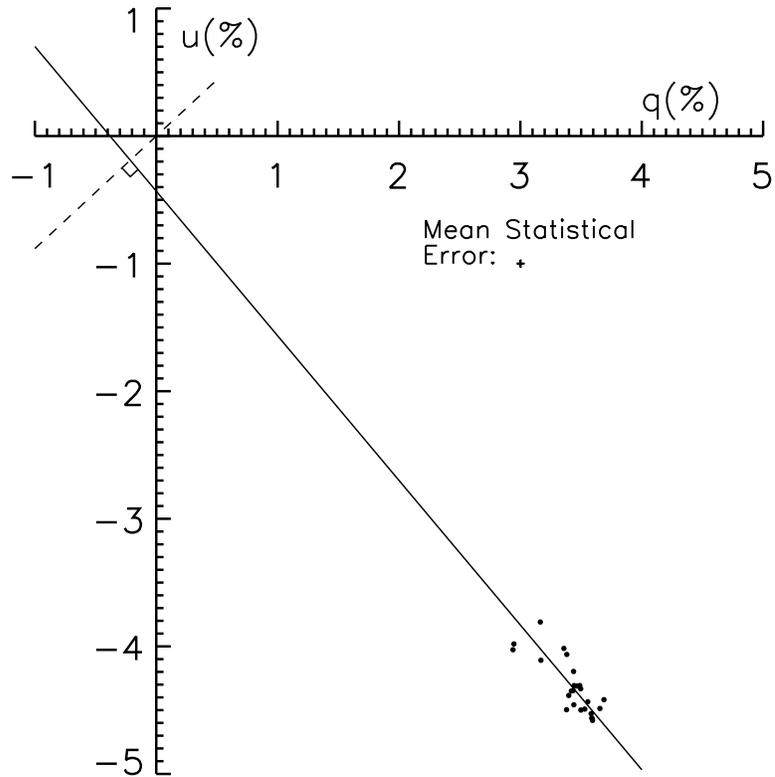}
}
}
\end{center}
\caption{Polarization data for SN~1999gi obtained on 2000 March 25, where each
point represents a bin of 100 \AA\ in each of the Stokes parameters.  The
best-fitting line, determined by a uniform-weight least-squares fit, is drawn
through the points, along with the line perpendicular to it that intersects the
origin.  The distance between the point of intersection of these two lines and
the origin reveals the component of wavelength-independent SN continuum
polarization in the direction perpendicular to the best-fitting line in the
$q$-$u$ plane, $p_{\perp} = 0.29 \pm 0.43\%$. Note that the statistical error 
of each point is insufficient to explain the scatter about the best-fitting 
line, which is likely produced by the effect of low-contrast line features 
in the spectropolarimetry (Fig.~\ref{fig:1}).
\label{fig:4}}
\end{figure}

The constancy of the polarization angle with wavelength across the optical
spectrum ($\theta$ never strays by more than $2^{\circ}$ from $154^{\circ}$;
see Fig. 8 of LF01) also casts doubt on large intrinsic SN polarization
since, unless the SN polarization P.A. is either parallel or
perpendicular\footnote{Recall that $\theta \equiv 1/2 \arctan (u/q)$, so that a
P.A. difference of $90^\circ$ corresponds to a $180^\circ$ orientation
difference in the $q$-$u$ plane.} to the ISP P.A., a somewhat unlikely chance
occurrence, significant variation in the polarization angle with wavelength
would be observed.  To quantify this assertion, we consider the representation
of polarization data in the $q$-$u$ plane (Fig.~\ref{fig:4}).  Unlike the
wavelength-independent nature of intrinsic SN continuum polarization, which
will produce a spherical cluster of points in the $q$-$u$ plane, a Serkowski
law wavelength-dependence of ISP should produce points that are spread out
along a given P.A.; due to noise, systematic uncertainty, and the effect of
line features, the points will not all lie exactly on a line, but will instead
have some scatter about it.  In the case of pure, undiluted ISP from a single
source, we would expect the best-fitting line to the points to intersect the
origin in the $q$-$u$ coordinate system, since this is the definition of having
a single P.A.  If another source possessing a significant polarization
component at $\pm 45^\circ$ to the ISP P.A. is contributing, then the
best-fitting line will not intersect the origin.  Under the simple paradigm of
a single ISP P.A. and wavelength-independent SN continuum polarization, in
fact, the amount by which the best-fitting line misses the origin reveals the
component of SN polarization oriented at $\pm 45^\circ$ to the ISP P.A. (i.e.,
$p_{\perp} \equiv \sqrt{q_{\circ}^2 + u_{\circ}^2}$, where $q_{\circ}$ and
$u_{\circ}$ are the points of intersection in the $q$-$u$ plane between the
line defined by the polarization data and the line perpendicular to it that
passes through the origin).

Figure~\ref{fig:4} shows the polarization data for SN~1999gi, where each of the
25 points represents a $100$ \AA\ bin in the $q$ and $u$ Stokes parameters.
The best-fitting line, determined by a uniform-weight least-squares fit, is
seen to miss the origin by only $p_{\perp} = 0.29\%$.  The statistical
uncertainty of the individual points are clearly insufficient to explain the
scatter about the best-fitting line, which is likely produced by the effect of
low-contrast line features in the spectropolarimetry (Fig.~\ref{fig:1}).
Therefore, to determine the uncertainty in the value of $p_{\perp}$, we
employed the technique of bootstrap resampling (Press et al. 1992), in which
new data sets of 25 points were created by randomly drawing data values from
the original 25 data points with repetition of values allowed.  We then
computed $p_\perp$ for 1000 data sets resampled in this way and found the $1\
\sigma$ scatter of the results to be $0.43\%$, which we adopt as the $1\
\sigma$ uncertainty on the value of $p_\perp$.  This convincingly demonstrates
that significant (i.e., more than $\sim 1.5\%$) wavelength-independent
intrinsic SN polarization must have a polarization angle either coincident
with, or perpendicular to, the ISP direction.

With this in mind, we now return to the curious observation that the observed
polarization of SN~1999gi {\it decreases} in the absorption troughs of strong
lines in the total flux spectrum (see Fig.~\ref{fig:1}), contrary to what is
expected theoretically and also to what has been observed in the
spectropolarimetry of other core-collapse SNe (e.g., SN~1987A [Jeffery et
al. 1991b, and references therein]; SN~1993J [Tran et al. 1997; H\"{o}flich et
al. 1996]; SN~1997ds [LF01]; SN~1999em [Leonard et al. 2001]; SN~2002ap
[Leonard et al. 2002a; Kawabata et al. 2002]).  The explanation for such atypical
behavior is straightforward, however, if ISP dominates the observed
polarization and the SN polarization has a P.A. that is oriented perpendicular
to the P.A. of the ISP (as we suspect may be the case from the earlier $q$-$u$
plane analysis).  That is, if the intrinsic SN polarization has a P.A. that
lies along the best-fitting line in Figure~\ref{fig:4}, but in the upper left
quadrant of the $q$-$u$ plane (i.e., at P.A. $\approx 64^\circ$, perpendicular
to the ISP P.A. of $\sim 154^\circ$), then {\it intrinsic} SN polarization
level increases would become polarization decreases in the {\it observed}
spectropolarimetry due to the ISP.  The unusual polarization decreases in the
line troughs may therefore be only an artifact of a much stronger
contribution from the ISP to the overall polarization level.

A final argument against large intrinsic SN polarization is the fact that no SN
II-P has ever been found to be intrinsically polarized at more than 1.5\%
(Wheeler 2000; LF01; Leonard et al. 2001). Indeed, in the detailed multi-epoch
spectropolarimetric study by Leonard et al. (2001) of SN~1999em, shown by L02
to be a spectral and photometric twin of SN~1999gi, the polarization does not
rise above $p_V = 0.5\%$ in the epochs studied (days 7, 40, 49, and 159 after
discovery).

The preceding discussion therefore presents a choice: either SN~1999gi has the
highest intrinsic polarization (by far) of any core-collapse SN yet observed, a
continuum polarization shape that is unlike the wavelength-independent
polarization expected from, and previously observed in, SNe II, and line
polarization features unlike those observed in any previous SN~II (i.e.,
polarization decreases in absorption troughs), or there is a large ISP
component. The simplest explanation, and one that also restores normalcy to all
of the peculiarities of the observed polarization, is that SN~1999gi possesses
only minimal intrinsic polarization, and that a large ISP dominates.  From the
analysis of the line polarization and continuum shape, as well as the
historical evidence against SNe~II-P possessing high intrinsic polarization, we
conclude that SN~1999gi is very likely to be intrinsically polarized in the
range $0.3 \leq p_{\rm SN} \leq 1.5\%$.  We also note that if the majority of
the intrinsic SN polarization is indeed oriented perpendicular to the P.A. of
the ISP as we suspect, then the true ISP level is {\it even greater} than the
level observed, due to the vector nature of polarization addition.

\subsection{Polarization Efficiency of Dust in NGC 3184}
\label{sec:polarizationefficiencyofdustinngc3184}

The most viable production mechanism for the bulk of the polarization observed
in SN~1999gi is therefore directional extinction by interstellar dust grains
along the l-o-s in NGC~3184.  Assuming the wavelength of maximum ISP to be
$\lambda_{\rm max} \approx 5100$ \AA\ (\S~\ref{sec:alternativestoisp};
Fig.~\ref{fig:1}) and adopting the empirical correlation between the wavelength
of maximum polarization and $R_V$ found by Whittet \& van Breda (1978), we
derive $R_V = 2.9$ for the dust polarizing the light of SN~1999gi (cf. LF01).
The theoretical upper bound on the dust polarization efficiency from
Equation~(\ref{eq:3}) is therefore
\begin{equation}
\frac{{\rm ISP}}{\Ebv} < 40\%\ {\rm mag}^{-1} .
\label{eq:6}
\end{equation}
\noindent If the dust has a polarization efficiency similar to typical MW dust,
of course, then the empirical upper limit given by Equation~(\ref{eq:4}) is
expected to hold.

% **** Figure 5 ****

\begin{figure}
\ssp
\begin{center}
\rotatebox{0}{
\scalebox{0.7}{
\plotone{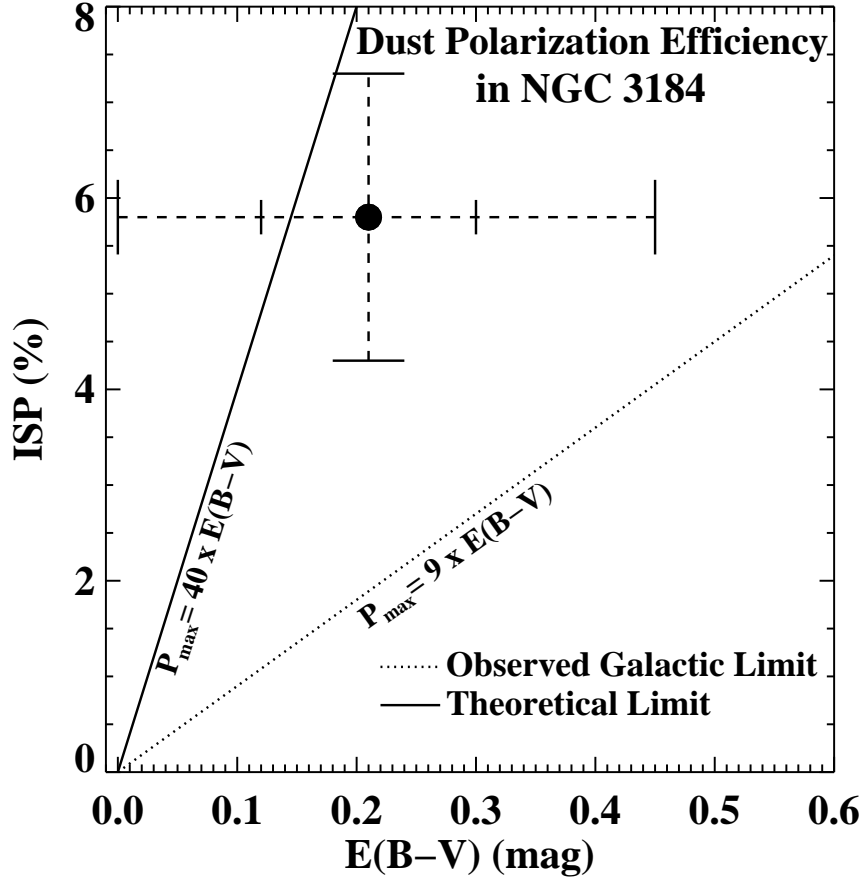}
}
}
\end{center}
\caption{The ISP and reddening along the l-o-s to SN~1999gi in NGC~3184, with
the observed polarization and best reddening estimate indicated by the {\it
filled circle}.  The $1\ \sigma$ reddening uncertainty derived by L02 is
indicated ({\it short error bars}), as are the lower and upper limits for both
the reddening and ISP ({\it long error bars}).  Also shown is the empirical
limit on the polarization efficiency of MW dust ({\it dotted line}; Serkowski
et al. 1975), as well as the theoretical upper limit for dust grains consisting
of completely aligned infinite dielectric cylinders ({\it solid line}; Whittet
1992).
\label{fig:5}}
\end{figure}

From the limits on the intrinsic SN polarization established in
\S~\ref{sec:alternativestoisp} as well as the null results for the Galactic
probe stars, we conclude that the polarization contributed by dust in NGC~3184
likely lies in the range $4.3\% \leq {\rm ISP} \leq 7.3\%$.  (We note that the
nature of the line trough polarization discussed in
\S~\ref{sec:alternativestoisp} actually favors restricting the ISP even
further, to values greater than 5.8\%; for the purpose of calculating the range
of {\it possible} polarization efficiency, however, it seems prudent to adopt
the most liberal limits.)  When combined with the reddening of $\Ebv = 0.21 \pm
0.09$ mag (L02), we derive a polarization efficiency\footnote{Due to the
non-Gaussian nature of the ISP probability distribution (i.e., it is flat
between the end points), the quoted values for the polarization efficiency and
uncertainty were derived through a Monte Carlo simulation.  We first randomly
selected 10,000 values from a Gaussian-weighted distribution for the reddening
($\EBV = 0.21 \pm 0.09$ mag) and a flat probability distribution for the ISP
($4.3 \leq {\rm ISP} \leq 7.3$).  From physical considerations (i.e., dust is
never observed to {\it deredden} light) we did not allow negative reddening. So
that the distribution of reddening values would remain symmetric about $\Ebv =
0.21$ mag, we also did not allow reddenings above $0.42$ mag.  This latter
restriction is quite justifiable as well from the upper limit of $\Ebv < 0.45$
mag derived by L02.  We therefore replaced any reddenings that were not within
the range $0.0 < \Ebv < 0.42$ mag with ones drawn randomly from the
Gaussian-weighted reddening distribution that do fall within the allowed range.
We then computed the polarization efficiency for the 10,000 sets of values.
The resulting probability distribution is not normally distributed, but rather
has positive skewness and a few extraordinarily large values resulting from
very small (but allowable) reddenings.  These outlier points bias both the mean
and the standard deviation to large, non-robust values (i.e., they cannot be
reproduced from one simulation to the next).  Our quoted result is therefore
the {\it median} of the probability distribution, and the lower and upper
values of the uncertainty mark the locations of the 16th and 84th percentile
points, respectively, of the simulated set of data (i.e., the $68\%$ confidence
bounds).} for the dust in NGC~3184 of $31^{+22}_{-9}\% {\rm\ mag}^{-1}$.
Figure~\ref{fig:5} displays the range of possible values for the ISP and
reddening for the dust in NGC~3184 along the l-o-s to SN~1999gi, with the
empirical and theoretical limits prescribed by Equations~(\ref{eq:4}) and
(\ref{eq:6}), respectively, indicated as well.  For lack of a clear physical
motivation for favoring a specific ISP value within the allowed range, we
nominally place it at ${\rm ISP} = 5.8\%$, the observed polarization value.

The range of possible ISP and reddening values shown in Figure~\ref{fig:5} is
inconsistent with the maximum observed Galactic polarization efficiency.  Our
nominal value is three times the empirical limit, and the upper end of the
range extends beyond the theoretical maximum given by Equation~(\ref{eq:6}).
Even the limiting values of $\Ebv = 0.45$ mag and ISP = 4.3\% produce a
polarization efficiency above (but just barely) the Galactic limit, and these
limiting values are thought to be quite unlikely.  Therefore, unless the
estimate of the reddening or the allowed range of the ISP are grossly in error,
this measurement is the {\it highest polarization efficiency yet confirmed for
a single sight line in either the MW or an external galaxy}.

\section{Discussion}
\label{sec:discussion}
\subsection{Galactic ISP Models}
\label{galacticispmodels}

What does the high polarization efficiency tell us about the dust in NGC~3184?
Does it suggest an unusually high degree of grain alignment?  An unusually
strong magnetic field?  Grain shapes that actually resemble infinite cylinders?
To gain insight we briefly discuss the various grain-alignment mechanisms that
have been proposed for the MW over the years.

First, it is clear that the Galactic magnetic field plays a leading
role in grain alignment, since the optical polarization position
angles in the MW of the most distant stars (e.g., Mathewson \& Ford 1970) are
generally found to parallel the Galactic magnetic field direction inferred from
other independent studies, largely at radio wavelengths (see, e.g., Heiles 1996
for a detailed comparison among the magnetic field directions derived from
starlight polarization, pulsar rotation measure data, and synchrotron
emission).  Similar results have also recently been
obtained using optical and infrared polarimetry of the {\it integrated}
starlight of a limited number of external galaxies as well, such as NGC~4565, a
normal spiral, in which the radio continuum (Sukumar \& Allen 1991), optical
polarimetry (Scarrott, Rolph, \& Semple 1990), and infrared polarimetry (Jones
1989) are all consistent with a magnetic field lying in the plane of the
galaxy.

The classic mechanism by Davis \& Greenstein (1951) achieves alignment through
paramagnetic relaxation of the thermally rotating grains in the presence of an
external magnetic field.  This mechanism, however, predicts a strong
correlation between the degree of alignment and magnetic field strength (e.g.,
Aannestad \& Purcell 1973; Johnson 1982), which is not observed (see, e.g.,
Heiles 1987 and references therein).  For this and other reasons, the classic
Davis-Greenstein mechanism is no longer seen as the most likely candidate for
alignment, although a modification of the basic idea is still favored (e.g.,
Jones 1996 and references therein).  

In recent years, the trend has been toward deemphasizing the role that
magnetic field {\it strength} plays in grain alignment and instead focusing on
its {\it direction}.  In fact, a very successful recent model by Jones, Klebe,
\& Dickey (1992) posits no correlation of alignment with magnetic field
strength at all (assuming, of course, that some nonzero field exists).  Rather,
the Jones et al. (1992) model assumes {\it complete alignment} of the large
aspherical component of the grain population at all times (i.e., independent of
the magnetic field strength), but with equal mixing of random and constant
directional components to the magnetic field.  That is, the model envisages a
uniform Galactic magnetic field pervading an interstellar medium that itself is
made up of individual cells containing superposed, randomly oriented magnetic
field vectors with a strength comparable to that of the Galactic field.  The
random components have the effect of causing the polarization angle of the net
field to vary from cell to cell, thereby reducing the overall polarization
efficiency of the ISM.

The Jones et al. (1992) model successfully explains the correlation of
polarization strength with extinction at 2.2 $\mu$m (i.e., the {\it K} band),
where it is possible to study stars that suffer over 100 mag of visual
extinction.  In addition to favoring equipartition between the random and
constant components of the magnetic field, a prediction that is also in good
agreement with radio observations (e.g., Heiles 1987, 1996), this model finds
the extent of the cells to be dictated not by size, but, rather, by extinction.
Specifically, the best agreement with the observations results when the random
magnetic field component decorrelates over a path length of $A_V \approx 1$
mag, a result found to be valid for both diffuse gas and dense clouds.  One
implication of this is that stars with $A_V \lesssim 1$ mag should have higher
polarization efficiency, on average, than those with greater extinction.  High
polarization efficiency, therefore, should prefer the very low-extinction
regime, where the effect of only a single cell polarizes the light.  

Within the context of the Jones et al. (1992) model, then, we might be able to
explain the high polarization efficiency, the surprising polarization angle
(i.e., that it is not aligned with the spiral arm; see, e.g., Scarrott et
al. 1990, 1991), and the nearly constant wavelength dependence of the
polarization angle if the bulk of the polarization of SN~1999gi is produced by
a single dust cell of highly aligned grains with $l \gg r$ whose orientation is
determined primarily by a local, randomly oriented magnetic field.  Even with
these specifications, however, it is possible that an unusually regular
magnetic field and/or different alignment mechanism may be required to explain
such high polarization efficiency.

\subsection{The Polarization Efficiency of Galactic Dust}
\label{sec:thepolarizationefficiencyofgalacticdust}

Although the low-extinction regime is only sparsely sampled by the data of Jones
et al. (1992; only 3 of the 99 stars studied have $A_V \lesssim 1$ mag,
and none is found to violate the empirical Galactic limit set by
Equation~[\ref{eq:4}]), it is instructive to review the low-extinction
polarization data that {\it are} available for Galactic stars, to see if the
extraordinary polarization efficiency indicated by SN~1999gi is unique.

The most complete collection of stellar polarization data is the agglomeration
of stellar polarization catalogs by Heiles (2000), which contains polarization
and, where available, reddening data for 9286 Galactic stars; in nearly all
cases, the quoted stellar reddening was derived in the original catalog by
comparing a star's intrinsic color (estimated from its spectral type) with the
observed color.  An investigation of the low-extinction regime of the Heiles
(2000) catalog is given by Fosalba et al. (2002), in which only stars with
listed degrees and angles of polarization, as well as positive extinctions, are
studied (several stars in the Heiles (2000) catalog have negative reddenings,
since these values were reported by the original catalogs from which Heiles'
data were drawn); this subsample constitutes $\sim$ 60\% of the stars in the
Heiles (2000) catalog.  When looking at the average polarization properties of
the stars, Fosalba et al. (2002) indeed find evidence for increasing average
polarization efficiency with decreasing reddening in the reddening range $\Ebv
< 1.0$ mag: $P_{\rm avg}/P_{\rm max} = 0.39 \Ebv^{-0.2}$, where $P_{\rm avg}$
is the average value of the polarization and $P_{\rm max}$ is the empirical
limit given in Equation~(\ref{eq:4}) for each reddening.  Although a few dozen
stars are found to exceed the Serkowski et al. (1975) limit, Fosalba et
al. (2002) pass over these violations without comment, perhaps due to the
inherent uncertainties in the techniques used to gauge stellar reddenings
(e.g., Teerikorpi 1990), a difficulty that has added caution in the
interpretation of polarization efficiency deduced by other researchers as well
(e.g., Berdyugin, Sn\aa re, \& Teerikorpi 1995).  An additional challenge when
interpreting polarization data in the very low-extinction regime is the bias
toward high polarizations that results from the positive-definite definition of
polarization (i.e., $p = \sqrt{q^2 + u^2}$) as well as the positive skewness
exhibited by its probability distribution (e.g., Miller et al. 1988; see
Leonard et al. 2001 for a thorough discussion of these issues).

To more securely investigate possible violations of Equation~(\ref{eq:4}) at
low reddenings, the recent dust maps of Schlegel et al. (1998) offer a useful
alternative to extinctions estimated by stellar color excess.  First, they
provide reddenings along Galactic lines-of-sight accurate to $\sim 16\%$.
Second, they represent the reddening produced by the {\it total} Galactic
dust column along a particular l-o-s, so that the value serves as a natural
upper bound on the reddening to any Galactic star.  Using the Schlegel et
al. (1998) reddening values will therefore result in {\it lower bounds} on the
derived polarization efficiencies of Galactic stars.  Thus, any star whose
polarization efficiency is found to significantly violate Equation~(\ref{eq:4})
when calculated in this way can be flagged as a potential exception to the
Serkowski et al. (1975) limit.  One point that must be kept in mind, however, is
that the Schlegel et al. (1998) maps have a resolution of $\sim 6\farcm1$
which, though capable of resolving quite small-scale features, certainly leaves
open the possibility of unresolved dust filaments.  Nonetheless, the exercise
is worthwhile as a preliminary investigation.

% **** Figure 6 ****

\begin{figure}
\ssp
\begin{center}
\rotatebox{0}{
\scalebox{0.7}{
\plotone{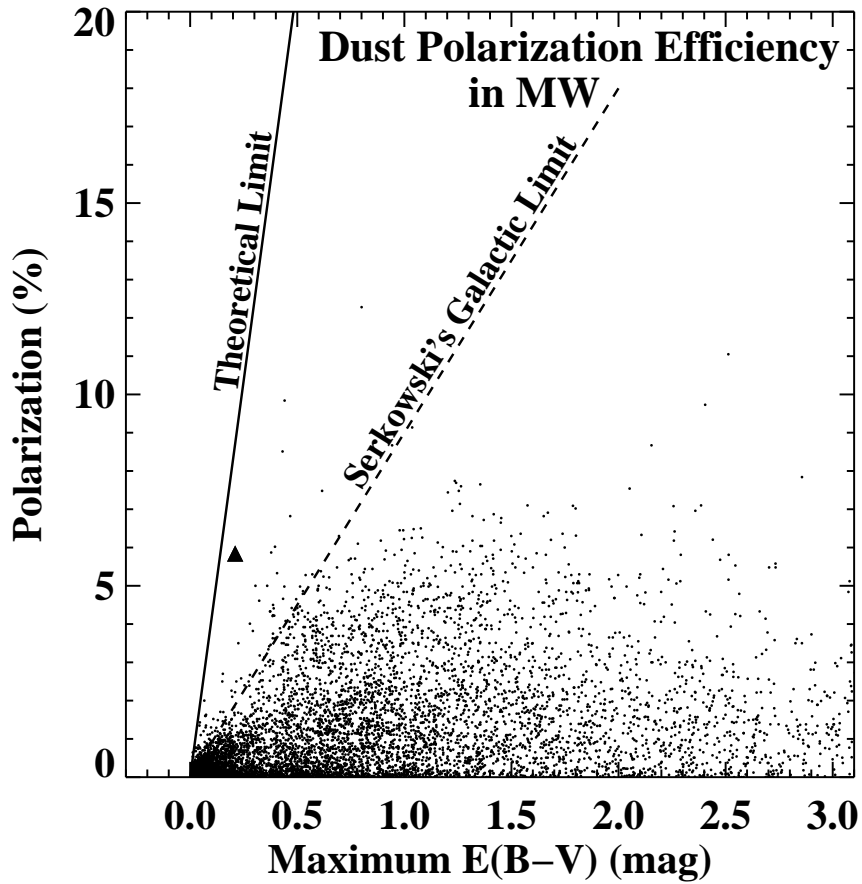}
}
}
\end{center}
\caption{Polarization of Galactic stars from the agglomeration catalog of
Heiles (2000) plotted against the Galactic reddening given by the dust maps of
Schlegel et al. (1998).  Since the Schlegel et al. (1998) reddenings probe the
entire Galactic dust column whereas the stars do not, the plotted reddenings 
represent upper bounds on the possible reddenings of these stars.  The
observed polarization and derived reddening of SN~1999gi is indicated by a 
{\it solid triangle}. 
\label{fig:6}}
\end{figure}

Figure~\ref{fig:6} shows the results of applying the Schlegel et al. (1998)
reddenings to the polarization data of Heiles et al. (2000) in the
low-extinction regime, and several stars clearly do violate the Serkowski et
al. (1975) limit.  Although we did not conduct a detailed investigation of all
of these stars, we did attempt to further investigate the most egregious
violators, for which the observed polarization exceeds the empirical limit by
more than 2\% (stars 3200, 3340, 3355, 3366, 3530, 3543, and 5997 in the Heiles
2000 list; note that the dust along the l-o-s to SN~1999gi likely violates the
empirical limit by $\gtrsim 2.5\%$).  Unfortunately, the source of the data for
all of these exceptional stars is an unpublished catalog by A. Goodman (see
Heiles 2000 and references therein) for which only positions and measured
polarizations, and no uncertainties or reddenings, are given.  This catalog
contains stars primarily in, near, or behind star-forming molecular clouds,
where the density of ${\rm H}_2$ molecules exceeds $10^2 - 10^3\ {\rm cm}^{-3}$;
such regions are noticeable as dark patches of extinction on optical
photographs and are therefore often referred to as ``dark clouds''.  From the
given positions we determined that all but the last star (Star 5997) lie in the
vicinity of the dark cloud L1755, a star-forming region in Ophiuchus, for which
additional information for two of the stars, numbers 3530 and 3543, are given
by Goodman et al. (1990) as $p = 5.87 \pm 0.12 \%$ and $p = 6.02 \pm 0.13\%$,
respectively.  The polarization values and stated uncertainties certainly place
these stars well above the empirical limit for their Schlegel et al. (1998)
reddenings (0.38 mag and 0.40 mag for Star 3530 and Star 3543,
respectively).\footnote{Note that the stars studied by Goodman et al. (1990)
were specifically chosen to be coincident, in projection (i.e., they are most
likely background stars), with the periphery of the extinction associated with
the dark clouds.  Such a selection criterion should naturally yield a sample
with low inferred reddening.}  At an estimated distance of $\sim 160$ pc to the
L1755 cloud (Bertiau 1958) the dust maps of Schlegel et al. (1998) are capable
of resolving structures of $\sim 0.3$ pc in size, which is sufficient to
resolve the L1755 cloud itself, estimated to have a physical size of $\sim 6$
pc $\times\ 0.7 $ pc (Goodman et al. 1990).  Certainly, though, it is
conceivable that unresolved substructure exists in such an environment, and it
is unfortunate that no specific extinction information exists for these stars.

Although the lack of specific reddening estimates precludes a definitive
statement, the possibility certainly exists that violations of the empirical
limit given by Equation~(\ref{eq:4}) do occur in the MW.  Additional studies of
the low-reddening regime are needed before more can be said about Galactic
violations of Equation~(\ref{eq:4}), however.  Until such studies are carried
out, though, our measurement of the polarization efficiency of dust in
NGC~3184, ${\rm ISP}/\Ebv = 31^{+22}_{-9}\% {\rm\ mag}^{-1}$, serves as the
highest polarization efficiency yet confirmed for a single sight line in either
the MW or an external galaxy.

\section{Conclusions}
\label{sec:conclusions}

The single spectropolarimetric epoch obtained for the Type II-P SN~1999gi by
LF01 indicates a very high observed polarization ($p_{\rm max} = 5.8\%$), with
a shape that is well characterized by a Serkowski law ISP curve.  The null
result obtained by spectropolarimetric observations of four Galactic stars near
to the l-o-s to SN~1999gi and sufficiently distant to fully sample Galactic
dust convincingly eliminates Galactic ISP as a significant contributor to the
observed polarization. It is also difficult to reconcile intrinsic SN
polarization, newly formed dust in the ejecta, or dust reflection by off-axis
dust blobs with the observed polarization characteristics.  We therefore
conclude that ISP produced by dust in NGC 3184 is largely responsible for the
observed polarization, and that it lies in the range $4.3\% \leq {\rm ISP} \leq
7.3\%$.  Since L02 find SN~1999gi to be only minimally reddened, $\Ebv = 0.21
\pm 0.09$ mag (with all reddening estimates supporting a firm upper limit of
$\EBV < 0.45$ mag, which was derived through the comparison between the
early-time continuum shape with that of an infinitely hot blackbody), we derive
a polarization efficiency for the dust along this l-o-s in NGC~3184 of
$31^{+22}_{-9}\% {\rm\ mag}^{-1}$.  Even by assuming the limiting reddening
value of $\Ebv = 0.45$ mag and the lower ISP limit of ISP = 4.3\%, the derived
polarization efficiency does not conform to the empirical Galactic limit of
${\rm ISP}/\EBV < 9\% {\rm\ mag}^{-1}$, and these limiting values are thought
to be extremely unlikely.  Therefore, unless the estimates for the ranges of
either the reddening or the ISP are grossly in error, this measurement is the
highest polarization efficiency yet confirmed for a single sight line in either
the MW or an external galaxy.

Although the polarization efficiency for dust along the l-o-s to SN~1999gi is
inconsistent with the empirically derived maximum observed Galactic
polarization efficiency, it can be accommodated by the maximum theoretical
efficiency derived from Mie scattering considerations (${\rm ISP}/ \EBV < 40\%\
{\rm mag}^{-1}$).  In addition, the dust grain polarization model of Jones et
al. (1992) predicts high polarization efficiency for lines of sight to sources
with low reddening ($A_V < 1$ mag).  From an examination of the Galactic
stellar polarization agglomeration catalog of Heiles (2000), we also find some
evidence for unusually high polarization efficiency for dust in dark clouds,
although the current lack of specific extinction information for these regions
precludes definitive conclusions.  Since the low-reddening regime has been only
sparsely sampled by previous studies, increased scrutiny of the polarimetry of
discrete Galactic and extragalactic sources with low reddening certainly seems
warranted to test whether the polarization efficiency inferred for dust along
the l-o-s to SN~1999gi in NGC~3184 is indeed common in the low-reddening
regime.

\acknowledgments

This research has made use of the SIMBAD database, operated at CDS, Strasbourg,
France, and was supported in part by NASA through the American Astronomical
Society's Small Research Grant Program.  Our work was also funded by NASA
grants GO-8648, GO-9114, and GO-9155 from the Space Telescope Science
Institute, which is operated by AURA, Inc., under NASA contract NAS 5-26555.
Additional funding was provided to A. V. F. by NASA/Chandra grant GO-0-1001C,
by NSF grant AST--9987438, by the Guggenheim Foundation, and by the Sylvia and
Jim Katzman Foundation.  KAIT was made possible by generous donations from Sun
Microsystems, Inc., the Hewlett-Packard Company, AutoScope Corporation, Lick
Observatory, the National Science Foundation, the University of California, and
the Katzman Foundation.

\end{document}